\documentclass[10pt, final, conference, compsoc, a4paper]{IEEEtran}
\usepackage{etoolbox}
\usepackage{xcolor}
\usepackage{todonotes}
\usepackage{amsthm}
\usepackage{amssymb,amsmath,amsthm,amsfonts}
\usepackage{physics}\usepackage{mathtools}
\usepackage{fancyhdr}

\DeclarePairedDelimiter\floor{\lfloor}{\rfloor}

\ifCLASSOPTIONcompsoc
  \usepackage[nocompress]{cite}
\else
  \usepackage{cite}
\fi
\ifCLASSINFOpdf
    \graphicspath{{./Figures/}}
    \DeclareGraphicsExtensions{.pdf,.jpeg,.png}
\else
\fi
\usepackage{algorithm}
\usepackage[noend]{algorithmic}

\newcommand{\R}{\mathbb{R}}

\newcommand{\xx}{\mathbf{x}}

\newcommand{\Psymb}{\mathbb{P}}

\newtheorem{theorem}{Theorem}[section]
\newtheorem{definition}{Definition}[theorem]

\begin{document}
%
\title{Improved Dynamics for the Maximum Common Subgraph Problem}
%
%
%
%

\author{
\IEEEauthorblockN{Davide Guidobene \IEEEauthorrefmark{1}\textsuperscript{1}}
\IEEEauthorblockA{Dipartimento di Scienze Ambientali,\\ Informatica e Statistica \\
Ca' Foscari University\\
Venice, Italy 30173}
\and
\IEEEauthorblockN{Guido Cera \IEEEauthorrefmark{1}\textsuperscript{2}}
\IEEEauthorblockA{Dipartimento di Scienze Ambientali,\\ Informatica e Statistica \\
Ca' Foscari University\\
Venice, Italy 30173}

}%

\IEEEtitleabstractindextext{%
\begin{abstract}

The Maximum Common Subgraph (MCS) problem plays a crucial role across various domains, bridging theoretical exploration and practical applications in fields like bioinformatics and social network analysis. Despite its wide applicability, MCS is notoriously challenging and is classified as an NP-Complete (NPC) problem. This study introduces new heuristics aimed at mitigating these challenges through the reformulation of the MCS problem as the Maximum Clique and its complement, the Maximum Independent Set. Our first heuristic leverages the Motzkin-Straus theorem to reformulate the Maximum Clique Problem as a constrained optimization problem, continuing the work of Pelillo \cite{Pelillo1999neurips} with replicator dynamics and introducing annealed imitation heuristics as in \cite{PavanPelillo2003} to improve chances of convergence to better local optima. The second technique applies heuristics drawn upon strategies for the Maximum Independent Set problem to efficiently reduce graph sizes \cite{Akiba2014}. This enables faster computation and, in many instances, yields near-optimal solutions. Furthermore we look at the implementation of both techniques in a single algorithm and find that it is a promising approach. Our techniques were tested on randomly generated Erdős-Rényi graph pairs. Results indicate the potential for application and substantial impact on future research directions.
\end{abstract}

\begin{IEEEkeywords}
Network theory, Simulated annealing
\end{IEEEkeywords}}

\maketitle

\footnotetext[1]{Current address: Computer Science Department, ETH Zurich}
\footnotetext[2]{Current address: Mathematics, Computer Science and Geoscience
Department, Università degli Studi di Trieste \\ \indent * Equal Contribution}

\IEEEdisplaynontitleabstractindextext

%
\IEEEpeerreviewmaketitle

\section{Introduction}
\label{sec:introduction}

%
%
%
%

\IEEEPARstart Finding the Maximum Common Subgraph (MCS) involves identifying the largest subgraph that is common between two graphs. This challenge is a fundamental issue in graph theory and computational science, recognized for its NP-Completeness \cite{Garey1979}. The MCS problem has profound implications across diverse applications such as molecular chemistry, pattern recognition, and network analysis, where it enables the detection of similar structures or patterns within complex networks. Similar to the Maximum Clique problem, the MCS problem is interconnected with various other combinatorial optimization challenges, including graph isomorphism, subgraph isomorphism, and network alignment. Solutions to the MCS problem range from exact algorithms, which are feasible for only small-scale problems due to their computational demands, to heuristic approaches that provide approximate solutions more efficiently for larger instances. These methods, while varying in their approach and accuracy, are crucial for tackling the vast array of applications that rely on graph similarity assessments, illustrating the balance between computational feasibility and solution optimality


\subsection{Maximum Clique Problem}
Finding a Maximum Clique means finding a complete subgraph with maximum cardinality. This problem is one of the most studied combinatorial problems and is also NPC \cite{Karp1972}. The problem is linked to other combinatorial optimization problems like, graph clustering, max-min diversity, optimal winner determination, graph vertex coloring, set packing and sum coloring. All this problems can be directly formulated as maximum clique problems or the maximum clique appears as a sub problem \cite{Wu2015}. For the maximum clique problem have been found both exact solutions and heuristics. Exact solution are mostly based on the branch-and-bound approach but they can be applied only on small instances because of their long computation times. Heuristics on the other hand, offer reasonable execution times but they don't guarantee an optimal solution. Both methods have different applications and serve different purposes. They are complementary approaches but they can also be combined in order to create more effective techniques.

\subsection{Maximum Independent Set Problem}
The Maximum Independent Set (MIS) problem is defined as follows, given a graph find a subgraph with maximum cardinality such that no vertex in the subgraph is adjacent to another vertex in the subgraph. It is a NP-hard problem and has many applications from finding the shortest route in a road network \cite{Kieritz2010} to computing the traverse order on the mesh borders for rendering \cite{Sander2008}. Like the other problems cited here exact solutions for the Maximum Independent set are not applicable on medium sized graphs or larger and heuristics are used instead. One of the most powerful heuristics used is called kernelization \cite{Akiba2014} and consists in reducing the size of the input to its most complex part, the kernel. In the case of MIS the kernel of a Graph $G$ is a graph $r(G)$ of smaller or equal size, obtained by applying a polynomial time algorithm to reduce the size of $G$ while preserving the relevant information to find a MIS of $G$. In practice kernelization is used as a preliminary step to facilitate other algorithms \cite{Butenko2009} or it can be the main portion of an algorithm if applied repeatedly \cite{Akiba2014}.

\subsection{Aim of this Paper}
This paper continues from the result obtained by Pelillo \cite{Pelillo1999neurips}. In his article Pelillo traces the Maximum Common Subgraph problem to the Maximum Clique problem. The results are comparable with state of the art deterministic combinatorial algorithms. Pelillo developed a framework in which the maximum clique problem is formulated as a undefined quadratic program. This framework enables a one-to-one correspondence between solutions of the quadratic program and solutions of the original problem. Finally the framework is promising ground for developing powerful heuristics. In this paper we present and test two approaches to enhance Pelillo's framework and we also demonstrate the soundness of using both techniques in series. The techniques are already known and established in graphs theory and in real world solutions. Some techniques have never been applied to the Maximum Clique problem directly and in such cases a clear connection to this problem from the original problem they were applied on is established.

\section{Techniques}
Note that in the following we only consider undirected graphs, unless otherwise specified.
First of all, we map the maximum common subgraph problem to the maximum clique problem, using the association graph.

\begin{theorem}
    Let $G'=(V', E'), G''=(V'', E'')$ two graphs of order n and m respectively (i.e$|V'| = n, |V''|=m$) and let $G=(V, E)$ be their association graphs.
    Then $C' \subseteq V'$ and $C'' \subseteq V''$ are a solution of the maximum common subgraph iff $C=\{(i, h) \in C'\times C''| h=\phi(i)\}$ is a maximum clique of $G$ (where $\phi: C' \to C''$ is the isomorphism between the subgraphs $G'[C'], G''[C'']$ induced by the two subsets of nodes).
\end{theorem}

Now, we can apply Motzkin-Straus theorem to turn the max-clique problem we are currently facing into a quadratic optimization problem.

\begin{definition}
    We denote with $\Delta_n := \{\xx \in \R^n | \forall i=1, ..., n (\xx_i \geq 0) \land \sum_{i=1}^n \xx_i = 1\}$ the standard simplex of $\R^n$.
\end{definition}

\begin{definition}
    Let $C\subseteq V$ a subset vertices of some graph $G$. We denote with $\xx^C \in \Delta_n$ its characteristic vector defined as
    \begin{equation*}
        \xx_i^C := \left\{
        \begin{array}{ll}
		\frac{1}{|C|}  & \mbox{if } i \in C \\
		0 & \mbox{otherwise}
        \end{array} \right. 
    \end{equation*}
    and we call $C$ the \textbf{support} of vector $\xx^C$.
\end{definition}

\begin{theorem}[\textbf{Motzkin-Straus}]
    Let $A \in \R^{N\times N}$ be the adjacency matrix of graph $G=(V, E)$ and let the function $f:\R^N \to \R$ be defined as $f(\xx):=\xx^TA\xx$. Then $C \subseteq V$ is a maximum clique of $G$ iff its characteristic vector $\xx^C$ is a global maximizer of $f$ over $\Delta_N$.
\end{theorem}

An interesting variation of Motzkin-Struas theorem has been proposed by Bomze \cite{Bomze1997} in 1997, which presents some interesting properties.


\begin{theorem}[\textbf{Bomze\cite{Pelillo1999neurips, Bomze1997}}]
    Let $\hat{A}:= A + \frac{1}{2}I_N$, where $A \in \R^{N\times N}$ is the adjacency matrix of graph $G=(V, E)$ and let $\hat{f}(\xx) := \xx^T\hat{A}\xx$. Then:
    \begin{itemize}
        \item $C \subseteq V$ is a maximum clique of $G$ iff its characteristic vector $\xx^C$ is a global maximizer of $\hat{f}$ over $\Delta_N$.
        \item $C \subseteq V$ is a maximal clique of $G$ iff its characteristic vector $\xx^C$ is a local maximizer of $\hat{f}$ over $\Delta_N$.
        \item All local maximizers of $\hat{f}$ over $\Delta_N$ are strict.
    \end{itemize}
\end{theorem}

\begin{figure}[!t]
    \centering
    \includegraphics[width=3.26in]{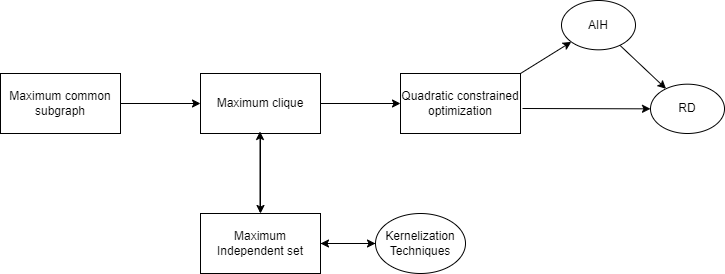}
    \caption{}
    \label{fig:scheme}
\end{figure}

\subsection{Replicator Dynamics (RD)}
Let's consider a game with a set of possible strategies $J=\{1, ..., n\}$, where the behavior of each player is predefined by the strategies they are assigned and cannot be chosen by that player.\\
We define the relative frequence of players that play strategy $i$ at time $t$ as $\xx_i(t) := \frac{N_i(t)}{N(t)}$, where $N(t)$ is the total number of players at time $t$ and $N_i(t)$ is the number of players that play strategy i at time $t$.\\
Given a payoff matrix, $W\in R^{n\times n}$ s.t. $(w_{ij} \geq 0) \forall i,j \in J$, the average payoff of strategy $i$ at time $t$ is given by $\pi_i(t) = \sum_{j=1}^n w_{ij} \xx_j(t)$.\\
In the discrete-time model, $N_i(t+1) = \pi_i(t) \xx_i(t)$, $N(t+1) = \sum_{i=1}^n N_i(t+1)$. Thereby defining the update rule:
\begin{equation}
    \label{RE}
    \xx_i(t+1) = \frac{N_i(t+1)}{N(t+1)} = \frac{\pi_i(t) \xx_i(t)}{\sum_{j=1}^n \pi_j(t) \xx_j(t)}
\end{equation}
On the other hand, the continuous-time time dynamical system is described by:
\begin{equation}
    \label{CRE}
    \frac{\partial}{\partial t} \xx_i(t) = \xx_i(t)\left(\pi_i(t) - \sum_{j=1}^n \pi_j(t) \xx_j(t) \right)
\end{equation}
These two equations are also known as \textit{replicator equations}.
\begin{theorem}[\textbf{Fundamental Theorem
of Natural Selection \cite{Pelillo1999neurips}}]
    Let $W \in R^{n\times n}$ be a matrix squared and symmetric s.t. $(w_{ij} \geq 0) \forall i,j \in J$ and let $F: \R^n \to \R$ be defined as $F\left(\xx(t)\right) := \xx(t)^TW\xx(t)$.
    Along any non-constant trajectory of replicator equations (\ref{RE}):
    \begin{itemize}
        \item $F$ is strictly increasing ($F\left(\xx(t+1)\right) > F\left(\xx(t)\right)$).
        \item $\xx(\cdot)$ converges to a (unique) stationary point ($\lim_{\tau \to \infty} \xx(\tau)$ is a stationary point).
        \item Assuming $diag(W) = \alpha I_n: \ \alpha \in (0,1)$, then $\xx(\tau)$ is asymptotically stable iff $\xx(\tau)$ is a strict (local) maximizer of $F$ over $\Delta_n$. \cite{PelilloTorsello2006}
    \end{itemize}   
\end{theorem}
This theorem suggests that replicator dynamics can be used to solve a constrained quadratic optimization problem over the simplex and this is exactly the approach implemented by Pelillo \cite{Pelillo1999neurips} using $W=\hat{A}$ as payoff matrix for the replicator dynamics in order to solve Bomze's optimization problem:
\begin{equation*}
    \begin{aligned}
        &\max \hat{f}(\xx) := \xx^T\hat{A}\xx\\
        &s.t.\ \xx \in \Delta_n
    \end{aligned}
\end{equation*}
Some practical considerations:
\begin{itemize}
    \item Without any better heuristics, it is advised to start replicator dynamics from the barycenter of the simplex $R(\Delta_n) := (\frac{1}{n}, ..., \frac{1}{n}) \in \Delta_n$.
    \item Empirical results indicate that it's better to first look for a stationary point using $W=A$ and then use that point to start replicator dynamics from there with $W=\hat{A}$.
    \item In order to verify that a stationary point $\xx^*$ is a (local) maximizer of $\hat{f}$ (while using $W=\hat{A}$) it suffices to check if $\xx^*$ is in the form of a set of the characteristic vector of a set of nodes $C \subseteq V$.
    \item If replicator dynamics stop at a stationary point $\xx^*$ that is not a (local) maximizer of $\hat{f}$ (while using $W=\hat{A}$), injecting gaussian noise will get the algorithm outside the undesired stationary point (this result can be justified using the Fundamental Theorem of Natural Selection that tells us $\xx^*$ is not stable).
    \item For the implementation of the algorithm we decided that the algorithm would stop when either $\|\xx_t - \xx_{t-1}\| \leq 10^{-6}$ or when it has reached the maximum number of iterations ($10^6$).
\end{itemize}

\subsection{Annealed Imitation Heuristics (AIH)}
\textit{Imitation dynamics (ID)} are a generalization of replicator dynamics described by the equation:
\begin{equation}
    \label{ID}
    \frac{\partial}{\partial t} \xx_i(t) = \xx_i(t)\left(\phi\left(\pi_i(t)\right) - \sum_{j=1}^n \phi\left(\pi_j(t)\right) \xx_j(t) \right)
\end{equation}
It can be easily noticed that (\ref{ID}) becomes equivalent to (\ref{CRE}) when $\phi$ is defined as the identity operator.\\
Let us also generalize the concept of support of a vector.
\begin{definition}
    A set of vertices $C \subseteq V$ is the support of a vector $\xx \in \Delta_n$ iff
    \begin{equation*}
        C = \sigma(\xx) := \left\{i \in V|\ x_i > 0 \right\}
    \end{equation*}
    Vice versa, let $C \subseteq V$ be a subset of vertices of some graph $G$ s.t. $|V|=n$. The \textbf{face} of simplex $\Delta_n$ corresponding to $C$ is defined as as:
    \begin{equation*}
        \Delta_{C} := \left\{\xx \in \Delta_n |\ \sigma(\xx)\subseteq C \right\}
    \end{equation*}
    and its \textbf{relative interior} as
    \begin{equation*}
        int\left(\Delta_{C}\right) := \left\{\xx \in \Delta_n |\ \sigma(\xx) = C \right\}
    \end{equation*}
\end{definition}
Given graph $G=(V, E)$ and its adjacency matrix $A$, let us now consider the following quadratic function
\begin{equation*}
    f_\alpha(\xx) = \xx^T (A-\alpha I_n) \xx
\end{equation*}
Let us define $\lambda_{max}(A)$ as the largest eigenvalue of matrix A (since A is symmetric, all its eigenvalues are real).
\begin{theorem}[\textbf{\hspace{1sp}\cite{PavanPelillo2003}}]
    If $\alpha > \lambda_{max}(A)$, then $f_\alpha$ is a function strictly convex in $\R^n$. Furthermore, its only solution $\xx$ belongs to $int(\Delta_C)$.
\end{theorem}
Let us denote by $A_C$, the submatrix of A formed by the rows and columns indexed by the elements of set $C \subseteq V$.
\begin{theorem}[\textbf{\hspace{1sp}\cite{PavanPelillo2003}}]
    Given $C \subset V$. If $\alpha > \lambda_{max}(A_C)$, $\nexists \xx \in int(\Delta_C):\ \xx$ is a local maximizer of $f_\alpha$.
\end{theorem}

Given a graph $G=(V, E)$ and a subset of vertices $C \subseteq V$, let us denote by $\gamma(C)$,
\begin{equation*}
    \gamma(C) := \max_{i\notin C} deg_C(i) - |C| + 1
\end{equation*}
where
\begin{equation*}
    deg_C(i) := \sum_{j \in C} A_{ij}
\end{equation*}

\begin{theorem}[\textbf{\hspace{1sp}\cite{PelilloTorsello2006}}]
    Given a graph $G=(V, E)$ and a subset of vertices $C \subseteq V$, $\gamma(C) = \lambda_{max}(A_C)$.
\end{theorem}

Given a graph $G=(V, E)$, let's define $\hat{\gamma}_m$:
\begin{equation*}
    \hat{\gamma}_m := 1 - (1-q)m - \sqrt{mq(1-q)\delta^\nu}
\end{equation*}
where $\delta \in (0,1)$, $\nu = \frac{|V| - m}{2}$ and $q$ is the density of $G$.

\begin{theorem}
[\textbf{\hspace{1sp}\cite{Bomze2002}\cite{PelilloTorsello2006}}]
Let $C \subseteq V$. If $|C| = m$, then
\begin{equation*}
    \Psymb(\gamma(C) > \hat{\gamma}_m) = 1 - \delta
\end{equation*}
\end{theorem}

Thanks to this interesting property, it makes sense to use $\hat{\gamma}_m$ as an approximation of $\gamma(C)$, choosing a small value for $delta$ (in our case $0.01$). Thus, given the clique number $c$, we want to choose $\alpha \in (\hat{\gamma}_{c-1}, \hat{\gamma}_c)$ so we can reasonably hope to avoid local optima. However, now the problem is we don't know $c$.

Pelillo and Torsello \cite{PelilloTorsello2006} showed in their paper a strategy to find an upper bound for the clique number of a graph.
\begin{theorem}[\textbf{\hspace{1sp}\cite{PelilloTorsello2006}}]
    Given a graph $G=(V,E)$, its clique number is at most $c_1 = \floor{\frac{\sqrt{8|E|+1}}{4} + \frac{1}{2}}$
\end{theorem}
Furthermore, in the case of the association graph, we know the maximum clique corresponds to the maximum common subgraph of the two original graphs $G_1=(V_1, E_1)$ and $G_2=(V_2, E_2)$. Thus, its clique number cannot exceed $c_2 = \min\{n_1, n_2\}$. Combining these results we get the following upper bound for the clique number of the association graph $G$:
\begin{equation*}
    c_{sup} = \min\{c_1, c_2\} = \min\{\floor{\frac{\sqrt{8|E|+1}}{4} + \frac{1}{2}}, n_1, n_2\}
\end{equation*}

So, our strategy consists in running these imitation heuristics using $alpha = \frac{\gamma_{m-1} + \gamma_m}{2}$ starting from $m=c_sup$ and then repeat the process iteratively reducing $m$ by one at every iteration. The idea is we know one of those $m$ will be the real clique number and thus, we hope that one of those heuristics will bring us closer to the global optima and thus, starting RD on Bomze's optimization problem, should give better results.

\subsection{Kernelization techniques}
Considering the complement $\Bar{G}$ of the association graph $G$, the sub-graph $C$ that we are searching is a maximum independent set (MIS) in $\Bar{G}$ instead of a maximum clique. This enables the application of multiple heuristic techniques to shrink the size of the problem; they are known as kernelization techniques. The following list contains a description of the techniques used. 
\begin{itemize}
    \item \textbf{Vertex Folding.} \cite{Chen2001} For a vertex $v$ of degree two whose neighbours $u$ and $w$ are not adjacent, either $v$ is in some MIS, or $u$ and $w$ are. The trio is contracted into a single vertex, if this is found to belong in a MIS then $u$ and $w$ belong to the MIS, otherwise $v$ is added.
    \item \textbf{Unconfined.} \cite{Xiao2013} Given $S=\{v\}, \ v\in V$, find $u \in V$ in the neighbourhood of $S$ $(N(S))$ such that $|N(u) \cap S|=1$. If there is no such vertex, $v$ is confined. If $N(u) \setminus N[S] = \emptyset$, then $v$ is unconfined. If $N(u) \setminus N[S]$ is a single vertex $w$, then add $w$ to $S$ and repeat the procedure; otherwise $v$ is confined. Unconfined vertexes can be removed from the graph, because there is always a MIS not containing such vertexes. 
    \item \textbf{Diamond.} \cite{Akiba2014} \cite{Iwata2016} This is an expansion of the Unconfined reduction. Using set $S$ built in the unconfined reduction for a vertex $v$ not unconfined. If there are non adjacent vertexes $u_{1}, u_{2}$ in $N(S)$ such that $N(u_{1}) \setminus N(S) = N(u_{2}) \setminus N(S) = {v_{1}, v_{2}}$, then $v$ can be removed from the graph since there is always a MIS without $v$.
    \item \textbf{Twin.} \cite{Xiao2013} Having two vertexes $u$ and $v$ of degree three such that $N(u)=N(v)$. If $G[N(u)]$ contains edges then $u$ and $v$ are in a MIS so we can remove $N(u)$ and $N(v)$. Otherwise contract $u$, $v$, $N(u)$ and $N(v)$ in a placeholder vertex $\xx$; if $\xx$ is in a MIS then $u$ and $v$ are in the MIS. This technique is a generalization of vertex folding.
    \item \textbf{Chang et al.'s Algorithm.} \cite{Chang2017} The LinearTime algorithm has complexity $O(m)$. It removes vertices of degree zero and one and applies a reduction rule on degree two nodes based on degree two paths. A degree two path is a path $P$ if every node of $P$ has degree two. A degree two path is maximal if it is not included in any other degree two path. For a maximal degree two path $P=(v_{1}, v_{2}, ..., v_{l})$ we call $v \notin P$ and $w \notin P$ the unique vertexes connected to $v_{1}$ and $v_{l}$ respectively. The technique distinguishes five cases depending on the length of the degree two path and its extremes.
    \begin{enumerate}
        \item $v=w$. There is a MIS that excludes $v$; therefore, we can remove $v$. In the following cases, we assume that $v \neq w$.
        \item $|P|$ is odd and $(v,w) \in E$. There exists a MIS that excludes $v$ and $w$; therefore, we can remove $v$ and $w$.
        \item $|P|$ is odd and $(v,w) \notin E$. There exists a MIS that excludes either $v_{1}$ or $w$ and includes half of the vertexes $\{v_{2}, ..., v_{l}\}$; therefore, we can remove $\{v_{2}, ..., v_{l}\}$ and add edge $(v_{1}, w)$.
        \item $|P|$ is even and $(v,w) \in E$. There exists a MIS that excludes either $v$ or $w$ and includes half of the vertices $\{v_{1}, ..., v_{l}\}$; thus, we can remove $\{v_{1}, ..., v_{l}\}$.
        \item $|P|$ is even and $(v,w) \notin E$. There exists a MIS that excludes either $v$ or $w$ and includes half of the vertexes $\{v_{1}, ..., v_{l}\}$; thus, we can remove $\{v_{1}, ..., v_{l}\}$ and add edge $(v, w)$.
    \end{enumerate}
\end{itemize}
Let $d(v)$ be the degree of node $v$. Let $V_{=0}, V_{=1}, V_{=2}, V_{\geq 3}$ be the sets of nodes in graph $G$ with degree zero, one, two and at least three, respectively. Let $S$ be an empty stack. Algorithm \ref{alg:linear_time} contains the exact procedure. Algortihm \ref{alg:procedures} contains auxiliary procedures used by algorithm \ref{alg:linear_time}.
\begin{algorithm}[t]
\caption{LinearTime}
\label{alg:linear_time}
\algsetup{linenosize=\small, linenodelimiter= }
\begin{algorithmic}[1]
\renewcommand{\algorithmicrequire}{\textbf{Input:}}
\renewcommand{\algorithmicensure}{\textbf{Output:}}
\renewcommand{\algorithmiccomment}[1]{/* #1 */}
\REQUIRE A graph $G=(V,E)$
\ENSURE  A maximal independent set $V_{=0}$ in $G$
\STATE \textit{Initialization()}:
\STATE Initialize a stack $S$ to be empty;
\WHILE{$V_{=1} \neq \emptyset$ or $V_{=2} \neq \emptyset$ or $V_{\geq 3} \neq \emptyset$}
    \IF{$V_{=1} \neq \emptyset$} 
        \STATE \textit{DegreeOne-Reduction()};
    \ELSIF{$V_{=2} \neq \emptyset$}
        \STATE \textit{DegreeTwo-Reduction()};
    \ELSE
        \STATE \textit{Inexact-Reduction()};
    \ENDIF
\ENDWHILE 
\STATE Iteratively pop a vertex $u$ from the stack $S$ , and add $u$ to $V_{=0}$ if none of its neighbors is in $V_{=0}$;
\STATE Extend $V_{=0}$ to be a maximal independent set;
\RETURN $V_{=0}$;
\end{algorithmic}
\end{algorithm}

\begin{algorithm}[t]
\caption{Auxiliary Procedures}
\label{alg:procedures}
\algsetup{linenosize=\small, linenodelimiter= }
\begin{algorithmic}[1]
\renewcommand{\algorithmiccomment}[1]{/* #1 */}


\item[] \textit{\textbf{Procedure} DegreeTwo-Reduction()}
\STATE $u \leftarrow$ a vertex in $V_{=2}$;
\STATE Find the maximal degree-two path/cycle $P=(v_{1}; $ $..., v_{l})$ containing $u$;
\IF{$P$ is a cycle}
    \STATE \textit{DeleteVertex($u$)};
\ELSE
    \STATE Let $v, w \notin P$ be the two vertices connected to $v_{1}, v_{l}$, respectively;
    \IF{$v=w$}
        \STATE \textit{DeleteVertex($v$)};
    \ELSIF{the number of vertices in $P$ is odd}
        \IF{there is an edge between $v$ and $w$ in $G$}
            \STATE \textit{DeleteVertex($v$)}; \textit{DeleteVertex($W$)};
        \ELSE
            \STATE Remove all vertices except $v_{1}$ of $P$ from $G$, remove all vertices of $P$ from $V_{=2}$, and add an edge between $v_{1}$ and $w$;
            \STATE Push vertices $v_{l}, ..., v_{2}$, obeying the order, into $S$;
        \ENDIF        
    \ELSE
        \STATE Remove all vertices of $P$ from $G$ and $V_{=2}$, and add an edge, if not exist, between $v$ and $w$;
        \STATE Push vertices $v_{l}, ..., v_{1}$, obeying the order, into $S$;
    \ENDIF
\ENDIF

\item[] \textit{\textbf{Procedure} Inexact-Reduction()}
\STATE Delete from $G$ the vertex $u$ with the highest degree by invoking \textit{DeleteVertex($u$)};

\textit{\textbf{Procedure} DeleteVertex($v$)}
\FOR{neighbour $w$ of $v$ in $G$}
    \STATE $d(w) \leftarrow d(w)-1$; \COMMENT{Remove edge $(v,w)$}
    \IF{$d(w)=2$}
        \STATE Remove $w$ from $V_{\geq 3}$ and add $w$ into $V_{=2}$;
    \ELSIF{$d(w)=1$}
        \STATE Remove $w$ from $V_{=2}$ and add $w$ into $V_{=1}$;
    \ELSIF{$d(w)=0$}
        \STATE Remove $w$ from $V_{=1}$ and add $w$ into $V_{=0}$;
    \ENDIF
\ENDFOR
\STATE Remove $v$ from $G, V_{=1}, V_{=2}, V_{\geq 3}$;

\textit{\textbf{Procedure} Contract($v,w$)}
\STATE Add edges, if not previously exist, between $w$ and neighbors of $v$;
\STATE Update $w$ to be in the correct set, $V_{=1}, V_{=2}, V_{\geq 3}$;
\STATE Remove $v$ from $G, V_{=1}, V_{=2}, V_{\geq 3}$;
\end{algorithmic}
\end{algorithm}

\subsection{Kernelization-AIH Pipeline}
Kernelization techniques can be added as a processing step to the AIH algorithm. The aim of this addition is to shrink as much as possible the graph to make the execution of AIH faster. Theoretically this is supported by the fact that, choosing the right kernelization techniques, they can be executed in linear time, while AIH algorithm is more complex as they use replicator dynamics and gradient descent. The kernelization algorithm cannot be the same used for its own experiment, because there most graphs get reduced completely and a solution is offered. The aim is only to remove as much vertexes as possible without losing information on a maximum clique in the graph, not to reach a solution; that is left to the AIH algorithm. For this reason the kernelization technique used is the Linear Time algorithm developed by Chang et al. \cite{Chang2017} with a slight adjustment. Chang et al.'s algorithm applies a procedure called by them \textit{Inexact-Reduction} which is very powerful and enables the algorithm to converge to a solution in most cases. This procedure also introduces most of the inaccuracy of the algorithm because it relies on very wide assumptions. Disabling this procedure and avoiding deleting vertexes found to be part of a maximum clique makes the Linear Time algorithm perfectly suited for our purpose, because it reduces the size of the graph but leaves room for AIH to operate with accuracy. Since the kernelization uses the complement of the association graph it is necessary to convert back the kernel before AIH is applied, this procedure is trivial [\figurename \ \ref{fig:scheme}]. 

\section{Experimental results}

\subsection{AIH vs RD}
Experiments are run on randomly generated Erdős-Rényi graphs of size 20 and different levels of expected connectivity. Every pair of graphs randomly generated by this algorithm is then used to build the corresponding association graph. Heuristics RD and AIH are then run to find a clique of the association graph as described in the previous section. Results in \figurename \ \ref{fig:aih} show AIH significantly improves over RD.


\subsection{Kernelization Techniques}

\begin{table}[t]
    \centering
    \caption{Kernel results}
    \begin{tabular}{|c|c|}%
        \hline
        \bfseries p & \bfseries Accuracy\\ 
        \hline
        0.01&0.95\\
        0.03&0.94\\
        0.05&0.93\\
        0.1&0.92\\
        0.2&0.90\\
        0.3&0.88\\
        0.4&0.88\\
        0.5&0.88\\
        0.6&0.88\\
        0.7&0.89\\
        0.8&0.90\\
        0.9&0.92\\
        0.95&0.94\\
        0.97&0.94\\
        0.99&0.95\\
        \hline
    \end{tabular}
    \label{tab:kernel_res}
\end{table}

The techniques are applied in the following order to maximise efficiency: LinearTime, VertexFolding, Twin, Unconfined and Diamond. The techniques have been tested on Erdős-Rényi random graphs of 100 vertexes and an expected connectivity $p$ taking the following values, 0.01, 0.03, 0.05, 0.95, 0.97, 0.99 and from 0.1 to 0.9 with 0.1 steps. For each value of $p$ 100 graphs were generated, then each was permuted to create isomorphic tuples; for a total of 1500 association graphs. These techniques have been tested on the Isomorphism problem, a more specific problem than the Maximum Common Subgraph in which the common subgraph is the size of the whole graph. The reduction times is longer on more sparse graphs. This is a direct consequence of the reduction rules used; many in fact only act if there are nodes with a low degree in the graph, one or two in several cases, and if there are very few such nodes or no nodes at all, the reductions perform little work, resulting in shorter times. In accordance with what has just been seen, accuracy increases precisely where the reduction rules have the greatest effect \figurename \ \ref{fig:acc}.

\subsection{Kernelization-AIH Pipeline}

\begin{table}[t]
    \centering
    \caption{RD vs AIH vs Pipeline}
    \begin{tabular}{|c|c|c|c|}%
        \hline
        \bfseries p & \multicolumn{3}{|c|}{\bfseries Size of found maximal common subgraph}\\
        \hline
        & \bfseries RD & \bfseries AIH & \bfseries kernel + AIH\\
        \hline
        0.1&13.54&\bfseries14.42&14.38\\
        0.2&10.81&11.69&\bfseries11.77\\
        0.3&9.37&10.22&\bfseries10.27\\
        0.4&8.71&\bfseries9.93&\bfseries9.93\\
        0.5&8.7&\bfseries9.96&\bfseries9.96\\
        0.6&8.58&\bfseries9.99&9.97\\
        0.7&9.15&\bfseries10.26&\bfseries10.26\\
        0.8&10.58&\bfseries11.58&11.54\\
        0.9&13.63&14.44&\bfseries14.47\\
        \hline
    \end{tabular}
    \label{tab:AIH_res}
\end{table}

\begin{figure*}[!t]
    \centering
    \begin{minipage}{0.45\textwidth}
        \centering
        \includegraphics[width=3.05in]{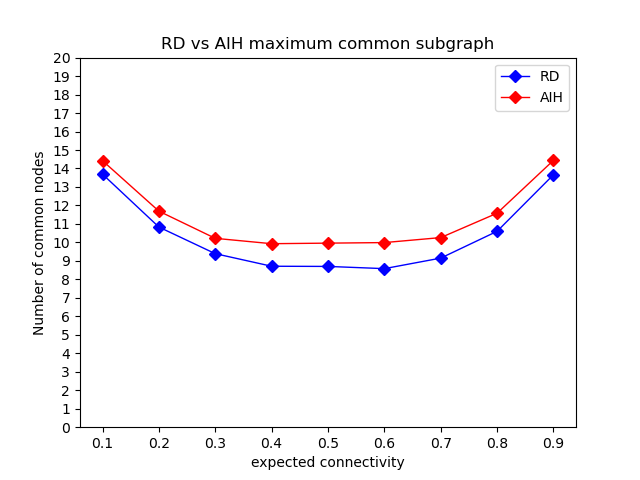}
        \caption{Size of maximum common subgraph found with RD and AIH heuristics comparison}
        \label{fig:aih}
    \end{minipage}\hfill
    \begin{minipage}{0.45\textwidth}    
        \centering
        \includegraphics[width=3.05in]{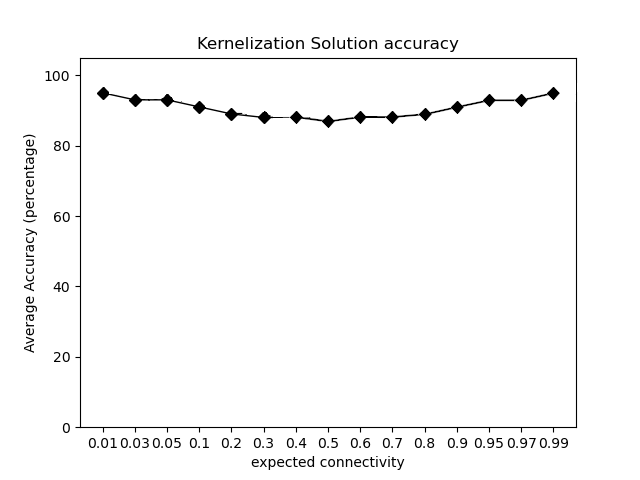}
        \caption{Average accuracy for graphs whose final kernel is empty. Numbers inside brackets refer to standard deviation.}
        \label{fig:acc}
    \end{minipage}
\end{figure*}
\begin{figure*}[!t]
    \centering
    \begin{minipage}{0.45\textwidth}
        \centering
        \includegraphics[width=3.05in]{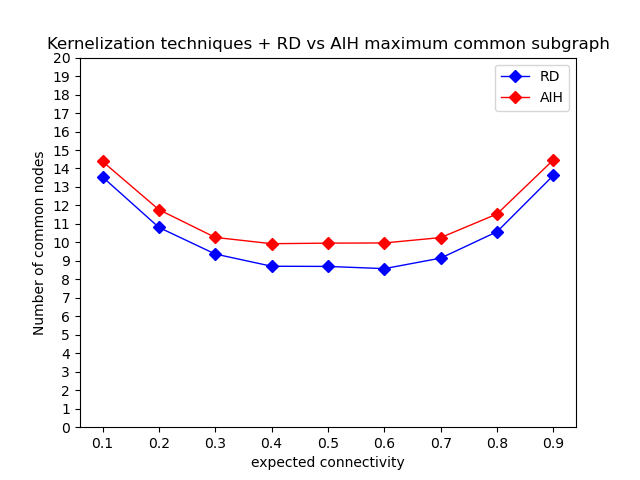}
        \caption{Size of maximum common subgraph found with kernelization + RD and kernelization + AIH heuristics comparison}
        \label{fig:k_aih}
    \end{minipage}\hfill
    \begin{minipage}{0.45\textwidth}    
        \centering
        \includegraphics[width=3.05in]{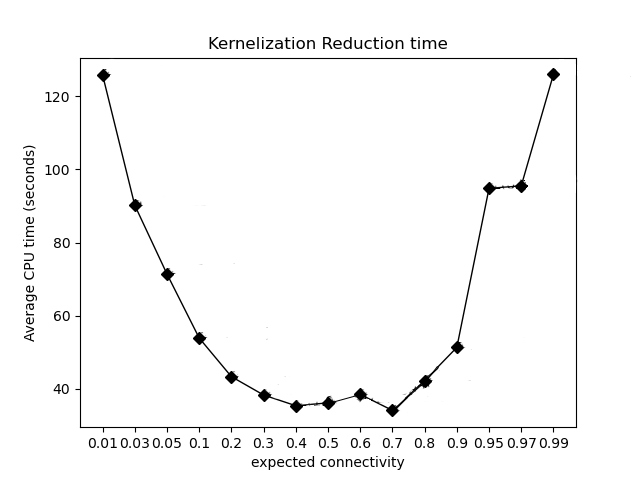}
        \caption{Average time to apply reduction techniques. Numbers inside brackets refer to standard deviation.}
        \label{fig:time}
    \end{minipage}
\end{figure*}

The data in \tablename \ \ref{tab:AIH_res} and \figurename \ \ref{fig:k_aih} show results of the pipeline being consistent with results of the AIH algorithm alone. The result that achieves the best accuracy between the two algorithms is highlighted in bold. \figurename \ \ref{fig:k_aih} shows similar improvement of kernelization+AIH over kernelization+RD compared to the one observed by pure AIH over RD in \figurename \ \ref{fig:aih}.
This validates our pipeline as a viable strategy since similar results have been reached starting from a smaller graph that retains the same information of the bigger, original graph. Due to the nature of how the experiments were conducted execution times are not really comparable between the single techniques and the pipeline but as explained in section 2 theory points to a favorable direction in this sense.

\section{Conclusion}
This paper presents techniques to enhance the framework developed by Pelillo and based on the findings of Motzkin and Strauss. 
Experimental results show that AIH finds significantly larger maximal common subgraphs than RD. Of course, this comes at the cost of higher computational time, as annealed imitation heuristics need to progressively adapt the starting point for the replicator dynamics. The second technique, kernelization, is one of the state-of-the-art techniques used for the Maximum Independent Set problem and, thanks to the framework used, it is easily applied to the Maximum Common Subgraph problem. Experimental results show the soundness of this approach, successfully shrinking graph size while retaining most of the relevant information and finding almost-optimal solutions. Kernelization techniques can be used in a pipeline with AIH techniques. With the proper adjustments, kernelization techniques can be used as a preprocessing step to simplify the association graph, therefore making AIH techniques faster, while keeping the maximum clique intact. However, experiments show that this simplification does not improve the size of the maximal common subgraph found by AIH.


%



\ifCLASSOPTIONcompsoc
  \section*{Acknowledgments}
\else
  \section*{Acknowledgment}
\fi
The work of this paper stems from Davide Guidobene's and Guido Cera's bachelor theses, which were made possible thanks to the tutoring of Full Professor Marcello Pelillo.




\ifCLASSOPTIONcaptionsoff
  \newpage
\fi


\clearpage
\begin{thebibliography}{99}

\bibitem{Pelillo1999neurips}
Marcello Pelillo; Replicator Equations, Maximal Cliques, and Graph Isomorphism. Neural Comput 1999; 11 (8): 1933–1955. doi: \\ https://doi.org/10.1162/089976699300016034

\bibitem{Bomze1997}Bomze, I.M. Evolution towards the Maximum Clique. Journal of Global Optimization 10, 143–164 (1997). \\ https://doi.org/10.1023/A:1008230200610

\bibitem{PelilloTorsello2006}Marcello Pelillo, Andrea Torsello; Payoff-Monotonic Game Dynamics and the Maximum Clique Problem. Neural Comput 2006; 18 (5): 1215–1258. doi: \\ https://doi.org/10.1162/neco.2006.18.5.1215

\bibitem{PavanPelillo2003}Pavan and Pelillo, "Dominant sets and hierarchical clustering," Proceedings Ninth IEEE International Conference on Computer Vision, Nice, France, 2003, pp. 362-369 vol.1, doi: \\ 10.1109/ICCV.2003.1238367.

\bibitem{Bomze2002}Immanuel M. Bomze, Marco Budinich, Marcello Pelillo, Claudio Rossi, Annealed replication: a new heuristic for the maximum clique problem, Discrete Applied Mathematics, Volume 121, Issues 1–3, 2002, Pages 27-49, ISSN 0166-218X, \\ https://doi.org/10.1016/S0166-218X(01)00233-5. \\ (www.sciencedirect.com/science/article/pii/S0166218X01002335)

\bibitem{Chen2001}Chen, J., Kanj, I. A., \& Jia, W. (2001). Vertex Cover: Further Observations and Further Improvements. Journal of Algorithms, 41(2), 280–301. https://doi.org/10.1006/JAGM.2001.1186

\bibitem{Xiao2013}Xiao, M., \& Nagamochi, H. (2013). Confining sets and avoiding bottleneck cases: A simple maximum independent set algorithm in degree-3 graphs. Theoretical Computer Science, 469, 92–104. https://doi.org/10.1016/j.tcs.2012.09.022

\bibitem{Akiba2014}Akiba, T., \& Iwata, Y. (2014). Branch-and-Reduce Exponential/FPT Algorithms in Practice: A Case Study of Vertex Cover. http://arxiv.org/abs/1411.2680

\bibitem{Iwata2016}Yoichi Iwata. (2016). Personal communication.

\bibitem{Chang2017}Chang, L., Li, W., \& Zhang, W. (2017). Computing a near-maximum independent set in linear time by Reducing-Peeling. Proceedings of the ACM SIGMOD International Conference on Management of Data, Part F127746, 1181–1196. \\ https://doi.org/10.1145/3035918.3035939

\bibitem{Karp1972}Karp, Richard M. 1972. ‘Reducibility among Combinatorial Problems’. Complexity of Computer Computations, 85–103. \\ https://doi.org/10.1007/978-1-4684-2001-2\_9

\bibitem{Wu2015}Wu, Qinghua, and Jin Kao Hao. 2015. ‘A Review on Algorithms for Maximum Clique Problems’. European Journal of Operational Research 242 (3): 693–709. \\ https://doi.org/10.1016/J.EJOR.2014.09.064

\bibitem{Kieritz2010}Kieritz, Tim, Dennis Luxen, Peter Sanders, and Christian Vetter. 2010. ‘Distributed Time-Dependent Contraction Hierarchies’. Lecture Notes in Computer Science (Including Subseries Lecture Notes in Artificial Intelligence and Lecture Notes in Bioinformatics) 6049 LNCS: 94–105. https://doi.org/10.1007/978-3-642-13193-6\_9

\bibitem{Sander2008}Sander, Pedro v., Diego Nehab, Eden Chlamtac, and Hugues Hoppe. 2008. ‘Efficient Traversal of Mesh Edges Using Adjacency Primitives’. ACM Transactions on Graphics (TOG) 27 (5): 144. https://doi.org/10.1145/1409060.1409097

\bibitem{Butenko2009}Butenko, Sergiy, Panos Pardalos, Ivan Sergienko, Vladimir Shylo, and Petro Stetsyuk. 2009. ‘Estimating the Size of Correcting Codes Using Extremal Graph Problems’. In Optimization: Structure and Applications, edited by Charles Pearce and Emma Hunt, 32:227–43. New York, \\ NY: Springer, New York. https://doi.org/10.1007/978-0-387-98096-6\_12/COVER

\bibitem{Garey1979}Garey, M., \& Johnson, D. (1990). Computers and Intractability; A Guide to the Theory of NP-Completeness. W. H. Freeman \& Co..



\end{thebibliography}
\end{document}